\documentclass[preprint,a4paper]{revtex4-1}

\usepackage[utf8]{inputenc}
\usepackage{graphicx,epsfig}
\def\be{\begin{equation}} \def\ee#1{\label{#1}\end{equation}}
\def\ba{\begin{array}}	  \def\ea{\end{array}}
\def\bea{\begin{eqnarray}}\def\eea{\end{eqnarray}}
\def\mc{\mathcal}         
\def\ci{\cite}  \def\ni{\noindent}
\def\ra{\rightarrow}  \def\DD{{_{\rm DD}}}  
\def\QCD{{\mbox{\scriptsize QCD}}} \def\F{{_{\rm F}}}
\def\P{{_{\rm P}}}   \def\tot{{_{\rm tot}}}  
\def\in{{_{\rm in}}} \def\el{{_{\rm el}}} \def\n{{_{\rm n}}}
\def\mc{\mathcal}

\begin{document}

\begin{center}
{\bf Inclusive processes in the modified Quark-Gluon String Model}\\
\vspace{0.2cm}
Mikhail N. Sergeenko\\ 
{\it Stepanov Institute of Physics of the National Academy 
of Sciences of Belarus,\\ BY-220072, Minsk, Belarus }\\
{\rm msergeen@gmail.com} 
\end{center}

\centerline{\bf Abstract} 
\vspace{0.2cm}
 Inclusive processes at high energies are studied in a non-perturbative 
approach in QCD using a modified Quark-Gluon String Model. 
 Theoretical and experimental aspects of diffraction dissociation are 
discussed.  
 In the calculations of cross sections, the parameters of complex 
nonlinear trajectories of Pomeranchuk and Reggeons are used. 
 Particular attention is paid to elastic and inelastic processes at 
LHC energies. 

\vspace{0.2cm}
\ni PACS: 11.10.St;12.39.Pn;12.40.Nn;12.40.Yx

\section*{Introduction}\label{sec:intro}
 It is agreed that the phenomena of hadron physics can be described 
in the framework of QCD~\ci{QCD03}. 
 There is a Lagrangian density of QCD, $\mc{L}_\QCD$, and in 
principle everything is derivable from it. 
 But the degrees of freedom in $\mc{L}_\QCD$ are quarks and gluons, 
not the hadrons we observe in nature. 
 This is why it is not so easy to make quantitative predictions for 
the hadron phenomena starting from $\mc{L}_\QCD$. 

 There are two areas where this has been successful: 
for short distances phenomena, where pQCD can be applied, and for 
hadron spectroscopy, where numerical or perturbative methods can be 
applied. 
 There is one more class of phenomena --- 
high-energy hadron-hadron ($hh$) collisions. 
 These reactions are classified into ``hard'' and ``soft'' ones. 
 In these processes, all energies and momentum transfers are assumed 
to be large. 
 The hard reactions can described using the perturbative QCD 
(pQCD)~\ci{PertQCD89}, the improved parton model.

 However, in hadronic collisions soft interactions are dominant and 
make the main contribution to the total cross sections. 
 These are characterized by large distances ($r\propto 1/\Lambda_{QCD}$) 
and the pQCD cannot be applied for their study. 
 On the other hand, Regge-pole theory is the main method for 
description of high-energy soft processes. 
 The Pomeron is the principal object in this approach, which is 
associated with cylinder-type diagrams of $1/N$ expansion of 
the scattering amplitude in QCD~\ci{tHoo74,Ven76}. 
 The $1/N$ expansion is a dynamical expansion and the speed of 
convergence depends on the kinematical region of studying process. 
 At very high energies many terms of the expansion (multi-pomeron 
exchanges) should be taken into account. 

 There are several successful Regge models: 
Dual Parton Model (DPM)~\ci{DPM81}, 
Quark-Gluon String Model (QGSM)~\ci{QGSM}, VENUS model~\ci{VEN93}, 
developed originally to describe soft hadronic interactions, and 
their modifications~\ci{PetRaBo92,LySe91,LySe96} 
to describe both soft and semihard hadronic reactions. 
 The QGSM is based on $1/N$ expansion in QCD and the theory of 
the supercritical Pomeron; it describes multiple production processes in 
$hh$ and hadron-nucleus collisions quite successfully. 
 There are the analytic~\ci{DPM81,QGSM} and 
the Monte-Carlo~\ci{AmeAll91} versions of this model. 
 However, the characteristics integrated over the hadron transverse 
momentum $p_\perp$ or at the average $p_\perp$ are usually considered 
within the framework of this model. 

 A modification of the QGSM to include the $p_\perp$-dependence was 
proposed in~\ci{LySe91,LySe96}. 
 To derive the dependence of the observed values on $p_\perp$ 
the mechanism of successive division of the internal transverse 
momentum $k_t$ between $2n$ quark-antiquark(diquark) chains 
(or $n$-pomeron exchanges) was suggested~\ci{LySe91}. 
 A strong dependence of the secondary hadrons on the number $n$ 
was observed. 

 Another modification of the QGSM was proposed in our work~\ci{MyPRD00}. 
 This two-component model was developed to analyze both soft and hard 
hadronic processes at high energies. 
 The longitudinal component is given by the string model and determines 
the behavior of the cross section on longitudinal variables. 
 The dependence on the transverse momentum is calculated on the basis 
of two-gluon Pomeron model in which the Pomeron is modeled as an exchange 
of two nonperturbative (NP) gluons whose propagator is finite at $q^2=0$. 
 Hard scattering of quarks on the ends of quark-gluon strings is 
calculated as a sequence of multi-Pomeron exchanges. 
 It was shown that the propagator which vanishes as $1/q^6$ or faster 
allows one to reproduce hard distributions of secondary hadrons. 
 The model was used to analyze the inclusive spectra of hadrons on 
the Feynman variable $x_\F$ and transverse momentum $p_\perp$ up to 
10 GeV/c in a wide energy interval. 

 In this paper, we use the modified QGSM to describe different hadron 
reactions at high energies.  
 We write down the invariant spectra in terms of topological cross 
sections and distributions of quarks (antiquark, diquarks) at the ends 
of quark-gluon strings.  
 In the calculations of quark disribution functions and fragmentation 
functions, 
the parameters of complex nonlinear Regge trajectories are used. 
 Theoretical and experimental aspects of diffraction dissociation 
are discussed.  

\section{Total cross sections in QGSM}
\label{totalCrSec}
 The $hh$ interactions at high energy are multiparticle processes 
and most events consist of the production of a large number of
particles with a small transverse momenta. 
 For such processes, the running coupling (which is not constant) 
is much too large for pQCD to be applicable. 
 Another feature of hadron interactions is that they involve 
multiple gluon exchanges. 
 The amplitude for single quark scattering on each other is not 
a sensible object~\ci{Nacht94}. 
 Sensible objects are the amplitudes for the scattering of hadrons 
that implies multiple gluon exchanges. 
 The problem is nonperturbative (NP) and should involve the
corresponding methods. 
 The first step in this way is to use the NP gluon propagator. 
 The second step should involve or effectively account for multiple 
gluon exchanges. 
 Therefore, alternative NP methods must be adopted.

 The theoretical descriptions of quantities like the total cross 
sections also should involve the NP QCD. 
 If the total cross section $\sigma_\tot(s)$ has a finite limit 
as $s\ra\infty$, then the $\sigma_\tot(s)$ in pure gluonic theory 
are NP objects and this conclusion is not changed if 
$\sigma_\tot(s)$ has a logarithmic behavior with $s$ for 
$s\ra\infty$.
 This can also be true in full QCD~\ci{Nacht94}.

 The total and the total inelastic cross sections 
of $hh$ interaction can be calculated in the framework of 
the ``eikonal approximation'' as follows~\ci{TerMar73}:
\be
\sigma_\tot(s) = \sum_{n=0}^{\infty}\sigma_\n(s),
\ee{sigt}
\be
\sigma_\in(s) =\sigma^\DD(s) +\sum_{n=1}^{\infty}\sigma_\n(s),
\ \ \ {\rm where}
\ee{sigi}
\be
\sigma_\n(s) = 
\frac{\sigma_\P(s)}{n\xi}\left(1-e^{-\xi}\sum_{k=0}^{n-1}
\frac{\xi^k}{k!}\right), \ \ \ \sigma_\P(s) = 8\pi\gamma_\P\zeta(s),
\ee{sign}
$\sigma^\DD(s)=\sigma_0(s)$ is the cross section of diffraction 
dissociation, $\sigma_\n(s)$ is the cross section for the production 
of the $n$ Pomeron chain, 
\be
\xi(s)=\frac{2C\gamma_\P}{R^2 +\alpha^\prime_\P(0)\ln(s/s_0)}\zeta(s),
\ \ \ \zeta(s)=\left(\frac s{s_0}\right)^\Delta,
\ee{xi}
$\Delta =\alpha_\P(0)-1\approx 0.08$, $\alpha_\P(0)$ is the intercept, 
$\alpha^\prime_\P(0)$ is the slope of the vacuum trajectory 
$\alpha_\P(t)$ at $t=0$, and $s$ is the squared total energy of 
colliding hadrons in the c.m.s., $s_0=1\,$GeV$^2$. 
 The parameter $C=1-\sigma^\DD/\sigma_\el$ accounts for the deviation 
from the eikonal approximation and $\sigma^\DD$ is the total cross 
section of diffraction dissociation. 
 The cross-section $\sigma_\P$ is the contribution of the supercritical 
Pomeron~\ci{LaNa87,QGSM} to the total cross section, 
parameters $\gamma_\P=3.45\,$GeV$^{-2}$, and $R^2=2.77\,$GeV$^{-2}$ 
(for $pp$ interaction) and determine the value of the Pomeron coupling 
with a hadron.

 The sum of the topological cross sections, $\sigma_\n$, yields the
total cross section for inclusive reactions (\ref{sigt})
and (\ref{sigi}). 
 This means that the quantities $\sigma_\n$ contain in integrated 
form all the information about the process under consideration, 
i.e., soft interactions, hard interactions, etc. 
 This fact can be used to determine the invariant inclusive cross 
section, $F(x_\F,\vec p_\perp)$, in the framework of the NP approach.

\section{Invariant spectrum in the modified QGSM}
\label{modQGSM}
 The main contribution to the processes of type $a+b \ra c+X$ 
is given by the cylinder-type graphs, cut in the $s$-channel. 
 The invariant hadron spectrum corresponding to these graphs 
is~\ci{LySe91,LySe96}
\be
F(x_\F,\vec p_\perp)\equiv E\frac{d\sigma }{d^3\vec p} =
\sum_{n=0}^\infty \sigma_\n(s)\phi_\n^h(x_\F,\vec p_\perp),
\ee{fxFpt}
where $\sigma_\n(s)$ are given by (\ref{sign}) and the functions 
$\phi_\n^h(x_\F,\vec p_\perp)$ describe the distributions of 
hadrons produced from the decay of the $2n$ strings. 
 The term with $n=0$ corresponds to diffraction dissociation. 
 The distributions $\phi_n^h(x_\F,\vec p_\perp)$ are written in 
terms of the light-cone variables~\ci{QGSM}. 

 Light-cone quantization of quantum field theory has emerged as 
a promising method for solving problems in the strong coupling 
regime.
 This method has a number of unique features. 
 It seems to be well suited to solving QCD and, contrary to other 
approaches, the relativistic wave functions transform trivially 
to a boosted frame. 
 Its language is close to experiment and phenomenology. 
 An important general feature of the behavior of the light-cone 
wave function is that each Fock component describes a system of 
free particles~\ci{BroAll92}. 

 The functions $\phi_n^h(x_\F,\vec p_\perp)$ in (\ref{fxFpt}) 
are written in the form
\be
\phi_n^h(x,\vec p_\perp) = \int_{x_+}^1dx_1\int_{x_-}^1dx_2
\psi_n^h(x,\vec p_\perp;x_1,x_2),\ \ \ {\rm where}
\ee{fixp}
\begin{eqnarray}
\psi_n^h(x,\vec p_\perp;x_1,x_2) = 
 {\cal F}_n^{qq}(x_+,\vec p_\perp;x_1)
\tilde{\cal F}_n^q(x_-,\vec p_\perp;x_2) \nonumber\\
+ {\cal F}_n^q(x_+,\vec p_\perp;x_1)
\tilde{\cal F}_n^{qq}(x_-,\vec p_\perp;x_2) \nonumber\\
+ 2(n-1){\cal F}_n^{q_{sea}}(x_+,\vec p_\perp;x_1) \nonumber
\tilde{\cal F}_n^{\bar q_{sea}}(x_-,\vec p_\perp;x_2).
\label{psi}  
\end{eqnarray}
 Here $x_{\pm} = \frac 12[(x_\perp^2+x_n^2)^{1/2}\pm x_n]$ are the
light-cone variables in the $n$ Pomeron chain, $x_\perp=2[(m_h^2+\vec
p_\perp^{\,2})/s]^{1/2}$, $x_1$, $x_2$ are the longitudinal
coordinates of quarks on the ends of the string, and $m_h$ is the
mass of the secondary hadron $h$.

 There are several methods to calculate the functions 
$\phi_n^h(x,\vec p_\perp)$~\ci{QGSM}. 
 In our approach the transverse momentum between the Pomeron 
showers is divided successively. 
 In this case the functions $\phi_n^h(x_\F,\vec p_\perp)$ are 
written in the form
\be
\phi_n^h(x,\vec p_\perp) = \sum_{k=1}^n\phi_1^h(x_k,\vec p_\perp),
\ee{fin}
where the longitudinal variable $x_k$ in the $k$th chain depends 
on the $k$ $x_k=x/(1-x_0)^{k-1}$, $x_0\simeq 0.35$.

 The functions ${\cal F}_n^\tau(x_+,\vec p_\perp;x_1)$, where 
$\tau=q, \bar q, qq$, $q_{sea}, \bar q_{sea}$, are the probabilities 
of production of the hadron $h$ from the fragmentation of the upper 
ends of the strings (beam fragmentation) and the functions 
$\tilde{\cal F}_n^\tau(x_-,\vec p_\perp;x_2)$ are the probabilities 
of production of the hadron $h$ from the fragmentation of the lower 
ends (target fragmentation). 
 These functions are represented by the convolutions 
\be
{\cal F}_n^\tau(x_{\pm },\vec p_\perp;x_{1,2}) = 
\int d^2\vec k_\perp\tilde f_n^{\tau,h}(x_{1,2},\vec k_\perp)
\tilde G_{\tau\rightarrow h}\left(\frac{x_\pm}
{x_{1,2}},\vec k_\perp;\vec p_\perp\right).
\ee{fnpm}

The quark distributions in the initial hadrons over $x$ and 
$k_\perp$ were taken in the factorized form  
\be
\tilde f_0^{\tau,h}(x,\vec k_\perp)=f_0^{\tau,h}(x)g_0(\vec k_\perp),
\ee{fg}
where $f_0^{\tau,h}(x)$ is the quark $x$ distribution in the initial
hadron, which has the form $f_0^{\tau,h}(x) = C\,x^{\alpha}
(1-x)^{\beta}$~\ci{QGSM}, and $g_0(\vec k_\perp)$ is 
quark $k_\perp$ distribution in initial hadron. 
 The factorized form (\ref{fg}) will also be true after the $n$ 
gluon exchange
\be
\tilde f_n^{\tau,h}(x,\vec k_\perp)=f_n^{\tau,h}(x)g_n(\vec k_\perp).
\ee{fgn}

The hadronization functions 
$\tilde G_{\tau\rightarrow h}(z,\vec k_\perp;\vec p_\perp)$ 
have been taken in the form~\ci{LySe91,LySe96}
\be
\tilde G_{\tau\rightarrow h}(z,\vec k_\perp;\vec p_\perp) =
G_{\tau\rightarrow h}(z)\tilde g(\tilde k_\perp),
\ee{gz}
where
\be
\tilde g(\tilde k_\perp) = \frac{\tilde{\gamma}}{\pi}
\exp(-\tilde{\gamma}\tilde k_\perp^2),\ \ \
\tilde k_\perp = \vec p_\perp - z\vec k_\perp,\ \ \
z = \frac{x_\pm}{x_{1,2}}.
\ee{gt}
 The functions $G_{\tau\rightarrow h}(z)$ on the right
hand side of (\ref{gz}) describe the $z$ dependence of 
the hadronization of the quark $\tau$ into the hadron $h$; 
these functions have the form~\ci{QGSM}
\be
G_{\tau\rightarrow h}(z) = a_h(1-z)^{\eta(\tau,h)+\lambda}f(z).
\ee{gzh} 
 Quantities $\eta(\tau,h)$ and $f(z)$ depend on the type of reaction.
 The functions $G_{\tau\rightarrow h}(z)$ and their parameters have
been described in detail in~\ci{QGSM}. 
 For the fragmentation $u\ra\pi^+$ (favored fragmentation) 
they are, for example, $\eta(u,\pi^+)=-\alpha_\rho(0)$ and $f(z)=1$, 
and for $d\ra\pi^+$ (unfavored fragmentation) we have
$\eta(u,\pi^+)=-\alpha_\rho(0)+1$ and $f(z)=1$. 
 Here $\alpha_\rho(0)$ is the intercept of the leading $\rho$ 
Regge trajectory corresponding to light $u$ and $d$ quarks. 

\section{The Pomeron}\label{Pomeron}
 The description of soft hadronic reactions at high energies is in
terms of Regge exchanges. 
 A Reggeon known as the Pomeron plays the dominant role in 
high-energy hadronic reactions. 
 There exist different approaches to investigating an object such 
as the Pomeron; ``soft'' and ``hard'' Pomeron are most popular 
of them. 
 The soft Pomeron is constructed from multi-peripheral hadronic 
exchanges and has intercept $\alpha_P(0) = 1$. 
 Because this is not compatible with the rising hadronic cross 
sections at high energies, the soft Pomeron was replaced by 
a soft supercritical Pomeron with an intercept 
$\alpha_P(0)>1$~\ci{LaNa87,QGSM}.

In the framework of QCD (in the scattering region), the Pomeron 
is understood as the exchange of two (or more) gluons~\ci{Low75,Nus75}. 
 The soft Pomeron can be considered as an exchange of two
NP gluons whose propagator does not show a pole at 
$q^2=0$~\ci{MyPRD00}.
 A connection between the nontrivial vacuum structure of QCD and 
soft high-energy reactions has been discussed in~\ci{LaNa87}. 
 It was argued that soft collisions should involve in an essential 
way the NP QCD. 
 A simple model of the vacuum developed by Landshoff and Nachtmann 
(LN)~\ci{LaNa87} explains the properties of the Pomeron observed 
in experiment. 
 In the framework of this approach, we reproduced the Pomeron 
coupling constant, $\beta_0^2=3.94$ GeV$^{-2}$, for the gluon 
condensate mass $M_c=0.993$\,GeV, and the correlation length 
$a=0.08$\,fm that fixes the parameters of the NP gluon propagator. 

\section{Diffraction dissociation}\label{DiffrDis}
 Inclusive processes we consider here, $a+b \ra c+X$, where $X$ 
denotes an unknown system of particles, can be characterized by 
the invariant distribution function (\ref{fxFpt}). 
 For unpolarized incident particles, the distribution (\ref{fxFpt}) 
does not depend on rotations around the beam axis. 
 It is thus a function of three essential variables. 
 One of  these is related to the center-of-mass (c.m.) energy 
$\sqrt{s}$, and for the remaining two variables, five sets are 
in general use. 
 Elastic scattering in hadron physics is a classic example of 
diffraction dissociation. 
 For our purposes we use the following two sets: ($x_F,\,p_\perp$), 
and ($t,\,s_X$), where $t=(p_a-p_c)^2$ is the squared momentum 
transfer, $s_X=(p_a + p_b - p_c)^2$ is the square of 
the invariant mass of  the unobserved  system (missing mass). 

 Transformation of the invariant function (\ref{fxFpt}) from one 
set of variables to another is performed using the corresponding 
Jacobian of the transformation of variables. 
 The invariant volume of the phase-space is transformed in 
our case as follows: 
\be 
\frac{d^3{\bf p}}E 
=\frac\pi{\lambda^{1/2}(s,\,m_a^2,\,m_b^2)}dt\,ds_X^2,
\ee{d3sE}
where $\lambda(s,\,m_a^2,\,m_b^2)=[s-(m_a+m_b)^2][s-(m_a-m_b)^2]$. 
 The invariant distribution function from here is 
\be 
f({\bf p};\,s) = E\frac{d^3\sigma}{d^3{\bf p}} 
=\frac 1\pi \lambda^{1/2}(s,\,m_a^2,\,m_b^2)\frac{d^2\sigma}{dt\,ds_X^2}.
\ee{Ed3sdt}

 Hard diffractive processes have been an active area of study at HERA 
and the Tevatron, and will continue to be so at the CERN LHC. 
 But one must be uneasy about all theoretical analyses if we do not 
even understand soft diffraction. 
 The data from the CERN ISR for soft diffraction dissociation, with 
the proton emerging with almost no energy loss, have been shown to be 
in good agreement with the triple-Regge description. 
 However, the higher-energy data from the CERN SPS collider do not 
show the increase of the cross section with energy expected from 
simple fits. 
 This has led certain authors to introduce unconventional features 
into the data analysis, which yield energy dependence in conflict 
with the standard notions of Regge theory. 
 In this paper, we show that this is unnecessary and that conventional 
Regge theory gives an adequate description of the data.

\section{Results and discussions}
 One Pomeron exchange plays the dominant role in high-energy 
hadronic reactions. 
 However, with the energy growth, multiple Pomeron exchanges 
begin to play an important role. 
 The modified QGSM represents a natural framework to calculate 
such processes. 

Invariant distribution $f(x)$ and $p_\perp$ distribution can be 
calculated with the help of the formulas
\be
f(x) = \int F(x,\vec p_\perp)d^2\vec p_\perp,
\ee{fxinv}
\be
\frac{d\sigma}{dp_\perp^2}=\pi\frac{\sqrt{s}}2\int F({x,\vec
p_\perp)\frac{dx}{E^*}},
\ee{dsdp}  
where $E^*$ is the energy of the hadron $h$ in the c.m.s. of 
the reaction and $F(x,\vec p_\perp)$ is given by (\ref{fxFpt}).
 Figre 1 shows the application of the model for the reaction 
$pp\ra B^+X$ at $\sqrt{s}=1.96$\,TeV
 More examples are in~\ci{MyPRD00,LKSB2009}. 

\begin{figure}[th]
\begin{center}
\includegraphics[scale=0.95]{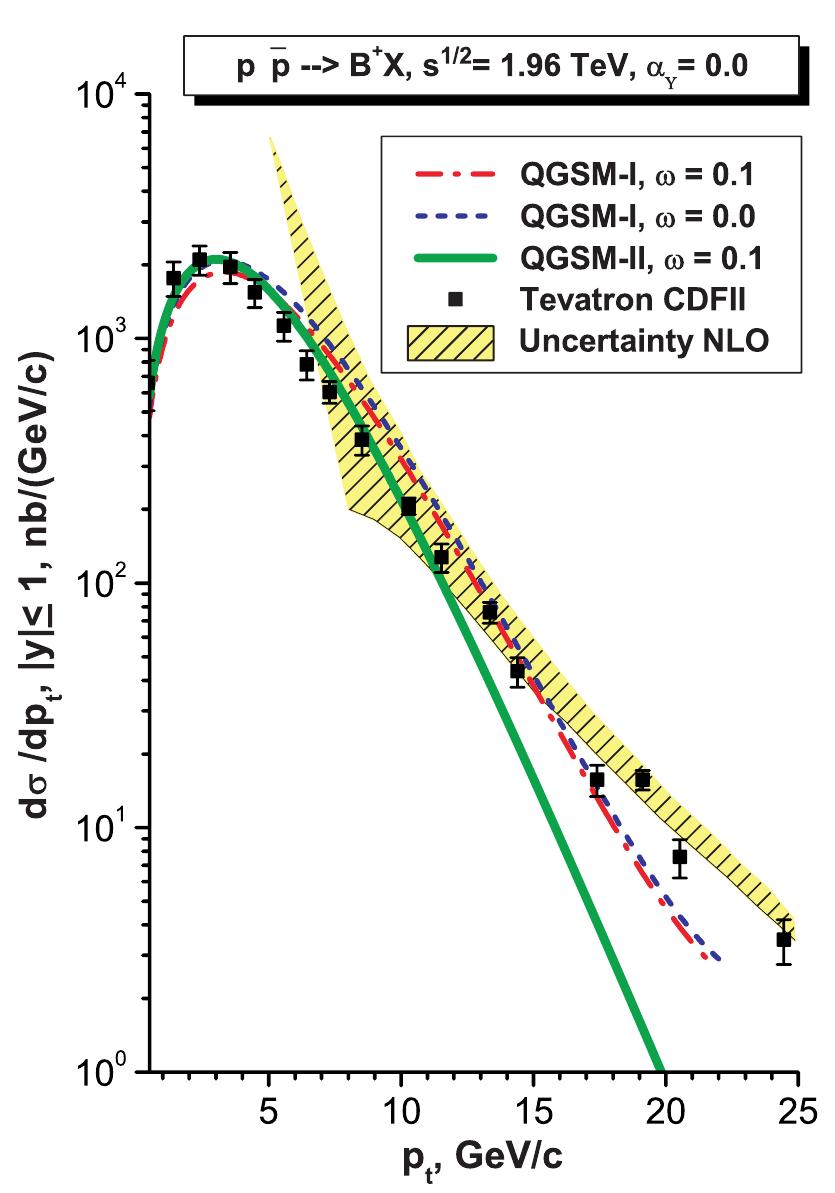}
\caption{{\small (Color online) The inclusive $p_\perp$ spectrum 
for $B^+$ mesons produced in the $pp$ collision at the Tevatron 
energy $\sqrt{s}=1.96$\,TeV obtained within the QGSM (solid and 
dashed lines) and within the NLO QCD~\ci{CDF03} (hatched regions); 
QGSM-I and QGSM-II correspond two sets of parameters in $p_\perp$ 
dependence~\ci{MyEPJC12}.}}
\label{fig1:CrBp196}
\end{center}
\end{figure}

\section*{Conclusion}\label{Concl}
 Hard collisions odd hadrons can be explained by the pQCD and models 
containing the semihard component. 
 The question that we have considered in this work is whether it 
is possible to apply the nonperturbative approach based on $1/N$ 
expansion of the scattering amplitude and the ``eikonal 
approximation'' to analyze the hard hadronic processes?

To deal with such a possibility we have used the fact that the first
term of the topological $1/N$ expansion corresponds to the
Pomeranchuk singularity and the next terms correspond to multi-Pomeron
exchanges. 
 We have used the fact that the total cross section and the total 
inelastic cross section of hadron-hadron interactions can be 
calculated in the framework of the ``eikonal approximation'' as 
the sum of the topological cross sections, $\sigma_n$, which give 
weights of the corresponding $n$ Pomeron-exchange diagrams. 
 The longitudinal component is given by the string model in which 
the $n$ Pomeron exchange is modeled by the forward scattering 
diagrams of the cylindrical type.
 According to Low and Nussinov we have modeled the Pomeron as the
two-gluon exchange, but we have modeled the exchange of two
nonperturbative gluons. 
 To calculate soft and hard hadronic processes, we have taken into
account the dependence of quark distributions in colliding hadrons
and the quark hadronization functions on the transverse momentum
$k_\perp$. 
 The color interaction of valence quarks, diquarks, and sea 
quarks (antiquarks) of the hadrons has been taken into account.
The essential parameters of the transverse component of the model, 
the value of the gluon condensate $M_c$ and the correlation length 
$a$, are both fixed by normalization conditions for the NP gluon 
propagator and can be calculated from the first principles.

{\bf Acknowledgments.} I would like to say a big thank you to my 
colleague and friend G.I. Lykasov and Z.M. Karpova for useful 
discussions and advices. 
 I especially thank Professor V.A. Petrov for the kind invitation 
to give a talk at the XXXVI International Workshop on High Energy 
Physics ``Strong Interactions: Experiment, Theory, Phenomenology, 
23--25 July 2024, Protvino, Russia. 
  
\bibliography{BiDaQM}
\bibliographystyle{apsrev4-1}

\end{document}